\documentclass[sigconf,authorversion,nonacm,screen]{acmart}

\AtBeginDocument{%
  \providecommand\BibTeX{{%
    \normalfont B\kern-0.5em{\scshape i\kern-0.25em b}\kern-0.8em\TeX}}}

\usepackage{soul}
\usepackage{balance}
\usepackage{tabularx}

\copyrightyear{2023}
\acmYear{2023}
\setcopyright{acmlicensed}\acmConference[CHI PLAY '23 Companion]{Companion Proceedings of the Annual Symposium on Computer-Human Interaction in Play}{October 10--13, 2023}{Stratford, ON, Canada}
\acmBooktitle{Companion Proceedings of the Annual Symposium on Computer-Human Interaction in Play (CHI PLAY '23 Companion), October 10--13, 2023, Stratford, ON, Canada}
\acmPrice{15.00}
\acmDOI{10.1145/3573382.3616099}
\acmISBN{979-8-4007-0029-3/23/10}

\begin{document}

\title{Kawaii Game Vocalics: A Preliminary Model}

\author{Katie Seaborn}
\email{seaborn.k.aa@m.titech.ac.jp}
\orcid{0000-0002-7812-9096}
\affiliation{%
  \institution{Tokyo Institute of Technology}
  \city{Tokyo}
  \country{Japan}
}

\author{Katja Rogers}
\orcid{0000-0002-5958-3576}
\affiliation{%
  \institution{University of Amsterdam}
  \city{Amsterdam}
  \country{The Netherlands}}
\email{k.s.rogers@uva.nl}

\author{Somang Nam}
\orcid{0000-0002-7785-9003}
\affiliation{%
  \institution{Toronto Metropolitan University}
  \city{Toronto}
  \country{Canada}}
\email{somang.nam@torontomu.ca}

\author{Miu Kojima}
\orcid{0009-0006-6122-6750}
\affiliation{%
  \institution{Tokyo Institute of Technology}
  \city{Tokyo}
  \country{Japan}
}
\email{kojima.m.ap@m.titech.ac.jp}

\renewcommand{\shortauthors}{Seaborn et al.}

\begin{abstract}
  Kawaii is the Japanese concept of cute++, a global export with local characteristics. Recent work has explored kawaii as a feature of user experience (UX) with social robots, virtual characters, and voice assistants, i.e., kawaii vocalics. Games have a long history of incorporating characters that use voice as a means of expressing kawaii. Nevertheless, no work to date has evaluated kawaii game voices or mapped out a model of kawaii game vocalics. In this work, we explored whether and how a model of kawaii vocalics maps onto game character voices. We conducted an online perceptions study (N=157) using 18 voices from kawaii characters in Japanese games. We replicated the results for computer voice and discovered nuanced relationships between gender and age, especially youthfulness, agelessness, gender ambiguity, and gender neutrality. We provide our initial model and advocate for future work on character visuals and within play contexts.
\end{abstract}

\ccsdesc[500]{Human-centered computing~Human computer interaction (HCI)}
\ccsdesc[500]{Applied computing~Computer games}
\ccsdesc[500]{Hardware~Sound-based input / output}
\ccsdesc[300]{Applied computing~Psychology}

\keywords{kawaii vocalics, kawaii, cuteness, vocalics, voice UX, game audio, user experience}

\received{20 February 2007}
\received[revised]{12 March 2009}
\received[accepted]{5 June 2009}

\maketitle

\section{Introduction}
Kawaii is the Japanese term for "cute," but it is also much more than that. Kawaii refers to impressions of cuteness, endearment, and even pity. As a psychosocial factor of perception and experience, kawaii has been linked to cognition, behaviour, and emotion \cite{nittono_behavioral_2010} within and beyond its native Japan \cite{nittono_cross-cultural_2021,fan2023cutenessfactor}. As a global phenomenon, kawaii has been adopted across a range of media around the world. Within the games space, kawaii can be attributed to series such as Pokémon, Animal Crossing, Kirby, and Super Mario Bros., as well as a wealth of other Japanese and Japanese-adjacent games\,\cite{schules_kawaii_2015,lin_beyond_2017,shibuya_male_nodate}. Perhaps the most famous kawaii game characters include Pikachu from Pokémon, the titular character of the Kirby series, and Toad from the Super Mario Bros. games. The appeal of these characters is near-universal, speaking to the transnational and cross-cultural characteristics of kawaii.

Many kawaii game characters are judged based on their appearance and behaviour: the visuals. Yet, character \emph{voice} may also be a key feature of their kawaii appeal and lasting value. From Pikachu's eponymous "pika pika" to Toad's vocal bursts, these vocalizations and speech patterns are instantly recognizable. Indeed, recent work in human-computer interaction (HCI) has pointed to the importance and dearth of work on kawaii sound \cite{zhang_exploring_2021}, especially for interactive characters like voice assistants \cite{seaborn_can_2023}. Moreover, the game context, with its complex audiovisual stimuli, may have special implications for kawaii vocalics, as there are many different attributes to consider. Voice can be direct intelligible speech or unintelligible gibberish (e.g., Simlish), presented in tandem with character animations. Characters can grunt or bark out lines (brief exclamations) or let out non-linguistic vocal bursts. Non-voicing sound techniques--type-writer or chalkboard sounds accompanying dialogue boxes\footnote{E.g., \url{https://www.youtube.com/watch?v=B3S9QCCMPRw}}--are also used. Voices and vocal expressions can appear to originate from within the game world (i.e., a visible character, often with speech animation), or present without a clear source in the game world (e.g., in menu interfaces or for characters not visible in the scene); while a matter of interpretation, this can be seen as a diegetic/non-diegetic distinction \cite{ekman2005meaningful}. Further, perception of vocalics (kawaii or otherwise) could interplay with dynamic and static visual character attributes, speech content (if applicable), and other auditory elements in the game (e.g., music or ambient noises). 
A model of kawaii game vocalics would offer a map of specific vocal attributes distinct from, but related to, other aspects of the game character known (or theorized) to invoke perceptions of kawaii, notably visual appearance and behaviour. This would increase our understanding of the relationship between vocal attributes or acoustical features and other well-studied game character features, and could be used by designers, researchers, critics, and even game fans who may be experimenting with new voice-generating technologies in gaming spaces on YouTube and Twitch\,\cite{lu2021more}. As yet, kawaii game character voices are virtually unexplored.

In this preliminary work, we propose an initial model of kawaii game vocalics: vocal features that lead to perceptions of kawaii in or through voices. Our goal was to motivate kawaii game vocalics as a new area of study in game user research by way of an initial model and empirical evidence of kawaii perceptions of in-game character voices.
To this end, we conducted an online perceptions study using a range of stimuli sourced from popular kawaii game characters in Japan. We asked two foundational research questions to elucidate the presence and perception of kawaii game vocalics. First, we asked RQ1: \emph{Are the voices of game characters deemed kawaii (implicitly by way of their visual appearance) also perceived as kawaii?} 
Second, we asked RQ2: \emph{Do these perceptions depend on the "girlish" qualities of the voices?} This question was motivated by two models of kawaii: the initial model of kawaii vocalics for computer voice \cite{seaborn_can_2023}, comprised of agedness (young), genderedness (girlish or ambiguous), language fluency, and anthropomorphism, and the original two-layer model of kawaii \cite{nittono_two-layer_2016}, comprised of visual cuteness and reactions to cuteness as a foundation alongside other perceptual attributes of the kawaii source, such as friendliness, harmlessness, and prettiness. 
Our contributions are (i) empirical evidence of kawaii voice perceptions in game characters based on audio stimuli alone; 
(ii) confirmation of the initial model of kawaii vocalics for game characters; and (iii) a data set of kawaii vocalics ratings for game characters. This work sets the stage for a new research agenda within the broader area of game sound and contributes to the games research community's understanding of a popular mode of expression in people and characters that, being associated with sexist and ageist attitudes \cite{shiokawa_cute_1999, seaborn_can_2023}, is still policed, discredited, and sidelined.

\section{Theoretical Background}

\subsection{A Psychosocial Model of Kawaii for Visuals and Voice}
Kawaii and cuteness as research phenomena have a relatively recent history with modern foundations\footnote{A history of kawaii is out of scope for this paper; we refer the reader to Shiokawa \cite{shiokawa_cute_1999} 
and Seaborn et al. \cite{seaborn_can_2023}, in ascending order of currency.}. Perhaps the first instance of "cute" being taken seriously as an object of study was by Lorenze in the earlier part of the last century\,\cite{lorenz_angeborenen_1943, nittono_behavioral_2010}. Lorenze proposed that perceptions of cuteness were linked to "baby schema" or Kindchenschema, referring to our innate reaction to the appearance of young animals, humans and otherwise: the roundness of features, large eyes, small mouths and noses, and so on. Nittono and colleagues \cite{nittono_behavioral_2010, nittono_power_2012, nittono_psychophysiological_2017, nittono_two-layer_2016,nittono_cross-cultural_2021,nittono_psychology_2022} later developed a two-layer model of kawaii as (i) \emph{positive emotions or affective responses} that stem from (ii) \emph{features of the socio-cultural environment}. Kawaii 
illustrates how phenomena that may originally relate to innate responses, i.e., Kindchenschema, are culturally influenced and socially mediated. The particular features of the  Japanese sociocultural context, notably amae, or a desire for acceptance, and chizimi shikou, or a preference for small things, highlight his interplay. Kawaii is now a decades-long cultural trend, firmly established in Japan but also embraced around the world: a cross-cultural phenomenon.

The two-layer model of kawaii is made up of a \emph{stimulus}, such as a kawaii object, which has \emph{attributes} deemed kawaii, such as Kindchenschema, smiles, and colours, that are \emph{perceived} in certain ways, such as cute, pretty, and gentle, leading to \emph{cognitive appraisals} of the stimuli as kawaii. These appraisals are revealed through \emph{emotional or affective responses}, which can \emph{manifest} subjectively (e.g., attitudes, thoughts), behaviourally, or physiologically. Thus we can chart a kawaii response to a given stimuli. Nevertheless, kawaii has essentially been explored as a visual phenomenon \cite{seaborn_can_2023}, with only a few exceptions on nonverbal sounds and melodies \cite{zhang_exploring_2021}, touch\,\cite{okada_can_2020}, behavioural conduct  \cite{lv_does_2021,marcus_cuteness_2017}, and voice \cite{lv_does_2021,seaborn_can_2023}. As Seaborn et al. \cite{seaborn_can_2023} point out, this leaves open the question of whether and how other modalities, notably sound, fit into the two-layer model of kawaii. We attempt to explore this within the context of games that feature kawaii voices.

The notion of kawaii has shifted over time, from its origins as pity for the weak to a non-threatening and endearing quality of people, non-human animals, and even objects \cite{nittono_two-layer_2016}. This has implications for games, as many characters--especially kawaii characters--are not human or even particularly humanlike, e.g., Pikachu, Kirby, Toad. Nevertheless, voice is an anthropomorphic, or at least biomorphic, feature not typically attributed to objects and generally associated with human-likeness \cite{seaborn_voice_2022}. Indeed, the initial model of kawaii vocalics provided evidence that human-likeness is key for kawaii perceptions in voice assistants. As such, we would expect that the game voices deemed most kawaii would also be the most humanlike and least artificial or machinelike. We thus hypothesized:

\begin{quote}
    H1. Perceptions of voice kawaiiness link to perceptions of low artificiality and high anthropomorphism.
\end{quote}

Kawaii has also been conceptualized as a "girlish" phenomenon or feature of expression. This has led to its dismissal as a serious topic by some \cite{inuhiko_kawaii_2006, nittono_two-layer_2016} and warnings of underlying sexism by others \cite{shiokawa_cute_1999, seaborn_can_2023}. At the same time, Nittono and colleagues have generally approached kawaii in a gender-neutral way, as well as provided some evidence of agelessness in the form of smiling elders deemed kawaii \cite{nittono_power_2012}. Seaborn et al. \cite{seaborn_can_2023}, focusing on voice phenomena, found preliminary evidence that gender ambiguity may be perceived as most kawaii for certain individuals. However, given that the voices deemed gender ambiguous were also deemed young, it is not clear to what extent gender, age, or the combination of the two explain perceptions of ambiguity. We attempt to further explore and distinguish the intersection of age and gender by including voices from human and non-human game characters. We thus hypothesized:

\begin{quote}
    H2. Perceptions of voice gender will link to kawaii by way of femininity, i.e., gendered feminine.
\begin{quote}
\end{quote}
    H3. Perceptions of voice age will link to kawaii by way of youthfulness, i.e., aged young.
\begin{quote}
\end{quote}
    H4. Perceptions of voice "girlishness" (feminine, young) will link to high kawaii ratings.
\end{quote}

\subsection{Theorizing Kawaii Game Vocalics within Game Audio Studies}

We approach kawaii game vocalics as a matter of perceiving auditory objects that are meaningful and may result in a variety of perceptions, cognitions, behaviours, and other reactions \cite{griffiths_what_2004}.
Voice in games is multifarious and can be situated within a range of game audio taxonomies. A common distinction concerns \emph{audio diegesis}  \cite{neumeyer2009diegetic,stilwell2007fantastical,liljedahl2011sound,jorgensen2010time}. For example, a voice's source of origin within the game world (tied to a visible character vs. disembodied) can influence whether it is considered diegetic or non-diegetic.
Liljedahl \cite{liljedahl2011sound} distinguishes between speech and dialogue, sound effects, and music. Some vocal audio in games would fit into the speech and dialogue category, but others (e.g., non-linguistic vocal bursts) could fit into the sound effects category, which, according to Friberg and Gärdenfors, refers to character, avatar, ambient, and ornamental sounds \cite{friberg2004audio}. All can involve vocal elements (e.g., general background noise of a crowd speaking---termed \emph{walla} if unintelligible). 
In a master's thesis, Holmes \cite[p.30]{holmes2021defining} categorized dialogue as main dialogue, reaction \& guidance lines (extending the former), ambient dialogue (background dialogue), walla (unintelligible ambient dialogue), combat barks (orders, commands, reactions b/w combatants), and emotes (``non-linguistic vocalisations such as screams, effort grunts, or breathing''), although it is unclear how this categorization was sourced or derived. 

Much work has been done on voice interaction in audio research (e.g., in games \cite{allison2018design}, voice assistants \cite{cambre2019one, sutton2019voice}, and robots \cite{mcginn2019can, torre2020trust}). Still, more work is needed on design guidelines for voice interface design \cite{murad2020designing}. Indeed, the vocal design of game characters and how voice impacts impression formation is a significant research gap.
Voices have certain characteristics that can convey emotions \cite{scherer2003vocal, murray1993toward}, impact trust in the speaker\,\cite{torre2020trust}, and influence perceptions of the personality \cite{brown1973perceptions,mcaleer2014how}. This impression formation is driven by characteristics like frequency or pitch \cite{mcaleer2014how}, harmonic-to-noise ratio \cite{mcaleer2014how}, and speaking rate or intonation \cite{brown1973perceptions}. 
However, we know very little about how this works for game characters, and how acoustical features in the voices of game characters contribute to players' impression of those characters.

Voice designers and voice actors in the games industry have a lot of training and experience in how to elicit certain impressions through vocalics. Indeed, such audio design and production roles are of growing importance\footnote{\url{https://blog.audiokinetic.com/en/a-speed-run-through-the-world-of-voice-design/}}. For example, professional voice (and motion capture) actors employ a full range of dramatic repertoire for character motion and 
vocal characteristics, such as intensity through physical exertion (e.g., by doing push-ups before or holding weights while recording\footnote{\url{https://www.youtube.com/watch?v=8sZMSBtSFrg}}). But non-linguistic speech (gibberish or vocal bursts) can also portray emotions (e.g., Bastion's melodic beeping in Overwatch \cite{overwatch}). 
Developing an empirical understanding of the vocal factors that can yield specific impressions could help (voice) audio practitioners hone their craft more quickly and contribute to basic knowledge.





\section{Methods}
We carried out an online user perceptions study using the research designs on visual \cite{nittono_power_2012} and voice-based \cite{baird_perception_2018} phenomena, aiming to replicate the kawaii vocalics study by Seaborn et al. \cite{seaborn_can_2023} with voice stimuli from game characters. Our protocol was registered before data collection on May 11\textsuperscript{th}, 2023 via OSF\footnote{\url{https://osf.io/7tr6b}} and approved by the research ethics board at the first author's institution.

\subsection{Participants and Recruitment}
Respondents (N=162) were recruited through Yahoo! Crowdsourcing Japan on June 12\textsuperscript{th}, 2023. Five incomplete responses were removed. The final set of participants (N=157) included women (n=76, one transgender), men (n=73), and none of another gender, while eight preferred not to say.  Most were aged 45-54 (n=49) or 35-44 (n=42), with some older (55-64 n=29, 65-74 n=11, 75+ n=2) and younger (18-34 n=16); eight preferred not to say. Most played games every day (n=49, 31\%), while others played games multiple times a week (n=38, 24\%), others never did (n=24, 15\%), some played once a week (n=14, 9\%) or once a month (n=13, 8\%), and others used to play games but stopped (n=11, 7\%), with eight preferring not to say. Participants were paid at $\sim$1200 yen per hour, equating to $\sim$300 yen for the study duration.

\subsection{Procedure}
Participants were given a link to a SurveyMonkey questionnaire containing the consent form, stimuli and rating scales, and demographic questions. After consenting on the first page, they were presented with the clips of game character voices, one voice per page (cf. Section \ref{3.3}). They were asked to rate these stimuli based on various vocal and social qualities related to kawaii (cf. Section \ref{3.4}). The stimuli were presented in a random order to avoid novelty and order effects \cite{schuman_questions_1996}. Demographics were collected on the final page to avoid priming effects \cite{head1988priming}. The study took $\sim$16 min.

\subsection{Materials} \label{3.3}
We selected 18 voices using a three-step process. Initially, we attempted to identify game character \emph{voices} deemed kawaii by experts and novices on magazines, websites, and social media. However, many focused on the voice actors, who were known beyond games and may have been judged based on their status as idols and celebrities in Japan. We also recognized that people may not associate kawaii with voice, given the lack of work on non-visual kawaii so far. We then instead identified game characters deemed kawaii \emph{in general}, using the same methods. Finally, we relied on our expertise to manually add characters that were missed. All authors decided on a final set of voice stimuli from all three steps that: (i) reflected a range of ages, genders, and vocal types for verifying our hypotheses; (ii) were diverse in terms of game of origin, i.e., representing a range of game types and particular game series or offerings; (iii) were similar in terms of game origin, e.g., Toad and Toadette; and (iv) where we could isolate the voice against any background sound or music. The first author recorded 2- to 20-minute voice clips from videos containing the original voices. From these, a diverse sample of vocal expressions were chosen and arranged into $\sim$8-second clips. The final set is in Table~\ref{tab:desc}.

\subsection{Instruments and Measures} \label{3.4}
All instruments used a 5-point Likert scale unless noted. The item order was randomized to curtail order effects \cite{schuman_questions_1996}. All items were translated into Japanese by a native speaker, and then back-translated into English and checked with an advanced speaker and native speaker. The full questionnaire was pilot tested with six native Japanese speakers.

\subsubsection{Kawaii Perceptions} In the absence of a validated measure, we used the one-item scale created by Seaborn et al.\,\cite{seaborn_can_2023}, which asks for an agreement rating of kawaii-ness. Given that it is a one-item scale, it was operationalized as a mean greater than 3.5 (skewed towards agreement on kawaii-ness) \emph{and} a median of 4 or above (nominal agreement). Marginal cases, where only one or the other of these metrics were met, are noted.

\subsubsection{Perceptions of Anthropomorphism, Artificiality, and Fluency} We used the one-item humanlikeness scale from Baird et al. \cite{baird_perception_2018}. Following Seaborn et al. \cite{seaborn_can_2023}, we divided the item into two items based on the poles (humanlike and artificial). We also added the item on language fluency, key for the kawaii vocalics model \cite{seaborn_can_2023}.

\subsubsection{Age Perceptions} Agedness was captured in a nominal scale: infant/baby (0-2 years), child (3-12 years), teenaged (13-19 years), adult (20-39 years), middle-aged (40-64 years), older adult (65+ years), and ageless.

\subsubsection{Gender Perceptions} Genderedness was captured in a nominal scale: feminine, masculine, aspects of both (reported as ambiguous), and neither (reported as neutral). Participants could also enter another option (free text).

\subsubsection{Demographics} We collected gender, age, education, and game use frequency.

\subsection{Data Analysis}
We used descriptive statistics to group voices by perceived age (baby, child, teen, adult, ageless) and gender (masculine, feminine, gender ambiguous, gender neutral).
When Shapiro-Wilks tests showed non-normal distributions, we used non-parametric statistics, e.g., 
Kendall's tau correlations. We also used Bonferroni corrections.

\section{Results}

Descriptive statistics are summarized in Table~\ref{tab:desc}. We now report on the hypotheses and full descriptive results.

\begin{table*}
  \caption{Descriptive statistics for perceptions of kawaii, age, and gender in game character voice, in order of kawaii ratings}
  \label{tab:desc}
  \begin{tabularx}{0.89\textwidth} {lllll}
    \toprule
    Character&Game or Series&Kawaiiness&Age Group&Gender Group\\
    \midrule
    
    Barbara & Genshin Impact \cite{gameGenshin} & Y: M=4.2, SD=.7, MD=4 & Child (MD=2, 60\%) & Fem. (MD=2, 97\%) \\
    Pikachu & The Pokémon series, e.g., \cite{pokemonlegendsarceus} & Y: M=4.2, SD=.7, MD=4 & Child (MD=2, 50\%) & Amb. (MD=2, 97\%)\\

    Edea & Bravely Default \cite{gameBD} & Y: M=3.9, SD=.7, MD=4 & Teen (MD=3, 48\%) & Fem. (MD=2, 87\%)\\
    Ayaka & Genshin Impact \cite{gameGenshin} & Y: M=3.9, SD=.9, MD=4 & Adult (MD=4, 87\%) & Fem. (MD=2, 99\%)\\
    QiQi & Genshin Impact \cite{gameGenshin} & Y: M=3.9, SD=.8, MD=4 & Child (MD=2, 58\%) & Fem. (MD=2, 94\%)\\
    Peach & The Super Mario series, e.g., \cite{supermariobros} & Y: M=3.7, SD=.8, MD=4 & Teen (MD=3, 43\%) & Fem. (MD=2, 92\%)\\
    
    Kirby & The Kirby series, e.g., \cite{kirbydeluxe} & Y: M=3.7, SD=.8, MD=4 & Child (MD=3, 69\%) & Amb. (MD=3, 51\%)\\
    Ashley & The Wario series, e.g., \cite{wariowaregold} & Y: M=3.7, SD=.8, MD=4 & Teen (MD=3, 68\%) & Fem. (MD=2, 96\%)\\
    Zelda & TLoZ series, e.g., \cite{tearsofthekingdom} & Y: M=3.6, SD=.8, SD=4 & Adult (MD=4, 63\%) & Fem. (MD=2, 98\%)\\
    Jigglypuff & The Pokémon series, e.g., \cite{pokemonredblue} & M: M=3.4, SD=1, MD=4 & Child (MD=2, 45\%) & Amb. (MD=3, 43\%)\\
    
    Toadette & The Super Mario series, e.g., \cite{supermariobros} & M: M=3.4, SD=.9, MD=4 & Child (MD=2, 59\%) & Fem. (MD=2, 55\%)\\
    Yoshi & The Super Mario series, e.g., \cite{supermariobros} & M: M=3.3, SD=1, MD=4 & Baby (MD=2, 57\%) & Amb. (MD=3, 53\%)\\
    
    Young Link & TLoZ series, e.g., \cite{majorasmask} & N: M=3.2, SD=.9, MD=3 & Child (MD=2, 62\%) & Masc. (MD=1, 55\%)\\
    Baby Bowser & The Super Mario series, e.g., \cite{supermariobros} & N: M=2.9, SD=1, MD=3 & Child (MD=2, 35\%) & Amb. (MD=3, 45\%)\\
    Toad & The Super Mario series, e.g., \cite{supermariobros} & N: M=2.9, SD=.9, MD=3 & Child (MD=2, 55\%) & Masc. (MD=1, 54\%\\
    Inkling Girl & Splatoon 3 \cite{gameSplatoon3} & N: M=2.7, SD=1, MD=3 & Child (MD=2, 41\%) & Amb. (MD=3, 41\%)\\
    Luma & Super Mario Galaxy \cite{gameSuperMarioGalaxy} & N: M=2.7, SD=1, MD=3 & Ageless (MD=7, 76\%) & Neu. (MD=4, 69\%)\\
    Shizue$^a$ & Animal Crossing, e.g., \cite{animalcrossingnewleaf} & N: M=2.6, SD=1, MD=3 & Ageless (MD=7, 70\%) & Neu. (MD=4, 57\%)\\

  \bottomrule
    \addlinespace
    \multicolumn{5}{@{}p{\dimexpr\linewidth}@{}}{\footnotesize Y: Yes. M: Marginal. N: No. TLoZ: The Legend of Zelda. Fem.: Feminine. Masc.: Masculine. Amb.: Ambiguous. Neu.: Neutral. $^a$Isabella.} 
\end{tabularx}
\end{table*}

\subsection{H1. Perceptions of voice kawaiiness link to perceptions of low artificiality and high anthropomorphism.}

A strong, negative correlation was found between Artificial and Humanlike ratings ($\tau$b = -.725, p < .05). A moderate positive correlation was found between Kawaii and Humanlike ratings ($\tau$b = .356, p < .05). A moderate, positive correlation was found between Humanlike and Fluent ratings ($\tau$b = .571, p <.05), and a moderate, negative correlation was found between Artificial and Fluent ratings ($\tau$b = -.480, p < .05). No others were found. As before \cite{seaborn_can_2023}, we can \textbf{accept the hypothesis that the most kawaii voices were also the most fluent, humanlike, and least artificial}.

\subsection{H2. Perceptions of voice gender will link to kawaii by way of femininity, i.e., gendered feminine.}

A Chi-Square test found a statistically significant association between Perceived Gender and Kawaii ratings across all voices, $\chi^2$(16, 2826) = 489.66, p < .05, $\phi$ = .208. Follow-up Chi-Square test of independence (Bonferroni corrected) showed relationships between Perceived Gender and Kawaii ratings for all categories: Feminine (M=3.8, SD=.8, MD=4, IQR=1), Ambiguous (M=3.4, SD=1, MD=4, IQR=1), Masculine (M=3.0, SD=.9, MD=3, IQR=2), Neutral (M=2.7, SD=1.1, MD=3, IQR=2), p < .05. No voice was classified as ‘Other’ across participants. 
A Kruskal-Wallis test indicated a significant difference by gender category, $\chi^2$(4) = 365.05, p < .05, with a Dunn’s test (Bonferroni corrected) revealing significant differences across all gender categories. As with computer voices \cite{seaborn_can_2023}, we can \textbf{accept the hypothesis that voices deemed feminine were perceived as the most kawaii}.

\subsection{H3. Perceptions of voice age will link to kawaii by way of youthfulness, i.e., aged young.}

A Chi-Square test found a significant relationship between Perceived Age and Kawaii ratings across all voices, $\chi^2$(24, 2826) = 404.52, p < .05, $\phi$ = .189. A Kruskal-Wallis test indicated a significant difference by age category, $\chi^2$(5) = 226.91, p < .05, with a Dunn’s test (Bonferroni corrected) revealing significant differences across all age categories except for Baby-Child and Teen-Adult: Teen (M=3.8, SD=.8, MD=4, IQR=1), Adult (M=3.7, SD=.9, MD=4, IQR=1), Child (M=3.5, SD=1.0, MD=4, IQR=1), Baby (M=3.4, SD=1.0, MD=4, IQR=1), and Ageless (M=2.7, SD=1.1, MD=3, IQR=2). As before \cite{seaborn_can_2023}, we can \textbf{accept the hypothesis that kawaii is an age-based phenomenon, linked to voice youthfulness}.

\subsection{H4. Perceptions of voice "girlishness" (feminine, young) will link to high kawaii ratings.}

Chi-Square tests found a significant relationship between Perceived Age and Kawaii rating for feminine voices, $\chi^2$(24, 1256) = 121.84, p < .05, $\phi$ = .156, gender ambiguous voices, $\chi^2$(24, 942) = 116.46, p < .05, $\phi$ = .176, and gender neutral voices, $\chi^2$(20, 314) = 37.99, p < .05, $\phi$ = .174, but not masculine voices. A Kruskal-Wallis test for voice and gender classifications revealed significant differences in Kawaii ratings, $\chi^2$(7) = 369.174, p < .05. A Dunn’s test (Bonferroni corrected) indicated this for all pairs except Feminine-Child and Feminine-Teen, Feminine-Child and Feminine-Adult, Feminine-Teen and Feminine-Adult, Ambiguous-Baby and Ambiguous-Child: Masculine-Child (M=3.0, SD=.9, MD=3, IQR=2), Feminine-Child (M=3.8, SD=.9, MD=4, IQR=1), Feminine-Teen (M=3.8, SD=.8, MD=4, IQR=1), Feminine-Adult (M=3.7, SD=.9, MD=4, IQR=1), Ambiguous-Baby (M=3.4, SD=1.0, MD=4, IQR=1), Ambiguous-Child (M=3.4, SD=1.1, MD=4, IQR=1), Neither-Ageless (M=2.7, SD=1.1, MD=3, IQR=2). We can \textbf{accept the hypothesis of kawaii as girlish but also gender ambiguous and gender neutral}, confirming previous work \cite{seaborn_can_2023} and demarcating \textbf{gender neutrality}. 


\section{Discussion}

\subsection{Kawaii Beyond Game Character Visual Appearance (RQ1)}
The results indicate that game character voices, even in the absence of other cues, especially visuals, can invoke a feeling of kawaii. In short, people can "hear cute." This gives weight to the importance of vocal design when creating game characters, especially if aiming for a kawaii impression. Yet, the level of kawaii varied across characters and voices deemed kawaii in ways that relate to the factors in the general and vocalics kawaii models, to which we now turn.

\subsection{Is Kawaii "Girlish"? Confirming and Advancing Models of Kawaii (RQ2)}

We found the expected patterns based on the kawaii vocalics model \cite{seaborn_can_2023} for this novel stimuli. We can confirm the extension to the two-factor model of kawaii for visual phenomena and social behaviour \cite{nittono_two-layer_2016} by including the social identity\,\cite{tajfel2004social} factors of gender and age, the computer agent factors of anthropomorphism and artificiality \cite{baird_perception_2018}, and the voice UX factors of vocal expressivity and fluency \cite{seaborn_voice_2022}. We also confirm the results on age and gender ambiguity found for computer voice \cite{seaborn_can_2023}. We also selected samples that allowed us to explore gender neutrality. While not well distinguished \cite{seaborn_neither_2022}, \emph{neutrality} speaks to an absence of gender, while \emph{ambiguity} speaks to the potential of pluralistic gender/ing. These \emph{gender liminal} perceptions may reflect generational shifts in attitudes towards the genderedness of kawaii: less about "girl" and "boy" varieties \cite{o2017fortheboys} and more about "individuality" \cite{shiokawa_cute_1999}. The results for anthropomorphism in combination with the factors in Table~\ref{tab:desc} also hint at patterns of liminiality for the non-human(oid) characters. The marginal cases of Jigglypuff (Child-Ambiguous), Toadette (Child-Feminine), and Yoshi (Baby-Ambiguous) juxtapose the low kawaii Luma (Ageless-Neutral) and Shizue (Ageless-Neutral). Toadette aside, all clips were vocal bursts or gibberish. Future work can explore biomorphism and vocal types against these factors in kawaii voice phenomena.

\subsection{Limitations}

We acknowledge several limitations. We conducted an online study with a relatively small set of vocal stimuli. 
More stimuli from characters deemed kawaii or not need to be explored. 
Also, we did not capture whether respondents knew the character belonging to the voice.
We thus cannot rule out the possibility that knowledge of the character and preexisting impressions of kawaii towards that character influenced results, which future work can tease out.

\subsection{Research Agenda}
As with the source phenomenon, kawaii game vocalics likely have repercussions for cognition, behaviour, and emotion that vary by game character and context, as well as by individual player. Here are some trajectories for future work:

\subsubsection{Game Visuals: Of Character and Context} An ongoing issue in research on game audio is that findings on audio perception captured \emph{outside} of games do not necessarily translate \emph{within} or \emph{across} games \cite{rogers2020potential}. Research on computer agents has long indicated that voice and body can and often do intersect in ways that may be influenced by other features of the interactive context \cite{seaborn_voice_2022}. The visual appearance of the character, including animations and behaviours, as well as the visual context, i.e., the visual game environment, will need to be teased out in terms of whether and how perceptions of kawaii in character vocalizations are influenced by relevant features of the game's visual modalities. 

\subsubsection{Towards a Taxonomy of Game Voice}

Game voice--of characters and beyond--is complex and multifaceted. One of the challenges we encountered was finding a rigorously developed framework by which to select and categorize voice phenomena for data analysis. This is a gap that needs addressing. The type of voice phenomena--from vocal bursts to gibberish--and the presence or absence of speech may affect kawaii perceptions. We relied on a preliminary taxonomy \cite{holmes2021defining} and our general knowledge of how voice phenomena is operationalized. The next step is to develop a game context-oriented taxonomy using rigorous methods like type-building analysis \cite{kuckartz2013qualitative}.


\subsubsection{The Kawaii Vocalics Data Set: An Open Science Endeavour}

As a "first" next step, we have started and offer an open data set. Like the ABOT (Anthropomorphic Robot Database) project \cite{phillips2018human}, we aim to provide a database of psychosocial factors attributed to specific voices. We provide our initial data set here: \url{https://bit.ly/kawaiigamevocalics}

\section{Conclusion}
Kawaii game vocalics brings a fresh perspective to the fields of game audio research and voice interaction.  Moreover, existing models of kawaii premised on visuals need to be expanded to include voice UX factors. Experimental work on a variety of non/humanoid characters and contexts will be needed--and there is a wealth of game material to draw on.

\begin{acks}
This work was supported by department funds at Tokyo Institute of Technology.
\end{acks}

\bibliographystyle{ACM-Reference-Format}
\balance
\bibliography{main.bbl}

\end{document}